\documentstyle[12pt,aaspp4]{article}
\newcommand{\be}{\begin{equation}}
\newcommand{\ee}{\end{equation}}
\newcommand{\bea}{\begin{eqnarray}}
\newcommand{\eea}{\end{eqnarray}}

\newcommand{\bef}{\begin{figure}}
\newcommand{\eef}{\end{figure}}

\def\spose#1{\hbox to 0pt{#1\hss}} 
\def\ltapprox{\mathrel{\spose{\lower 3pt\hbox{$\mathchar"218$}} 
 \raise 2.0pt\hbox{$\mathchar"13C$}}} 
\def\gtapprox{\mathrel{\spose{\lower 3pt\hbox{$\mathchar"218$}} 
 \raise 2.0pt\hbox{$\mathchar"13E$}}} 
\def\inapprox{\mathrel{\spose{\lower 3pt\hbox{$\mathchar"218$}} 
 \raise 2.0pt\hbox{$\mathchar"232$}}} 
 
 \slugcomment{Submitted to Astrophys. J. Letters}

\lefthead{Baertschiger, Joyce, Sylos Labini}
\righthead{Power-law correlation and discreteness  
in cosmological N-body simulations}

\received{2002 September 26}
\begin{document} 
 
\title{Power-law correlation and discreteness  
in cosmological N-body simulations}
\author{Thierry Baertschiger\altaffilmark{1,4}, 
Michael Joyce\altaffilmark{2,5,6} and  
Francesco Sylos Labini\altaffilmark{3,5,7}}

\altaffiltext{1}{thierry.baertschiger@physics.unige.ch}
\altaffiltext{2}{joyce@lpnhep.in2p3.fr}
\altaffiltext{3}{francesco.sylos-labini@th.u-psud.fr}

 \altaffiltext{4}{D\'epartement de Physique Th\'eorique, 
Universit\'e de Gen\`eve,
24 quai Ernest Ansermet, CH-1211 Gen\`eve 4, Switzerland} 

\altaffiltext{5}{Laboratoire de Physique Th\'eorique, 
Universit\'e Paris XI, B\^atiment 211, F-91405   
Orsay, France }

\altaffiltext{6}{Laboratoire de Physique Nucl\'eaire et de Hautes Energies 4, 
Place Jussieu, Tour 33 -Rez de chaus\'ee, 75252 PARIS Cedex 05 }

\altaffiltext{7}{INFM Sezione Roma1,        
		      Dip. di Fisica, Universit\`a ``La Sapienza'', 
		      P.le A. Moro, 2,  
        	      I-00185 Roma, Italy. }

\begin{abstract} 
We analyse with simple real-space statistics
the Virgo consortium's cosmological N-body simulations. Significant 
clustering rapidly develops well below the initial mean 
interparticle separation $\langle \Lambda_i \rangle$, where 
the gravitational force on a particle is dominated by that with
its nearest neighbours. A power-law behaviour in the two point 
correlation function emerges, which in the subsequent evolution 
is continuously amplified and shifted to larger scales, in a roughly 
self-similar manner. 
We conclude that the density fluctuations at the 
smallest scales due to the particle-like nature 
of the distribution being evolved are thus essential 
in the development of these correlations, and {\it not}
solely, as usually supposed, the very small 
continuous (fluid-like) fluctuations 
at scales larger than $\langle \Lambda_i \rangle$.
\end{abstract}

%\keywords{galaxies: general; galaxies: statistics; cosmology: 
%large-scale structure of the universe}

\setcounter{footnote}{0}

One of the central projects of contemporary cosmology is
to explain within a single framework the small fluctuations
observed in the temperature of the cosmic microwave background
radiation (CMBR) {\it and} the large fluctuations 
observed in the distribution of 
galaxies and other discrete objects.
The current paradigm (\cite{pee93,pad93})
for this framework is the cold dark matter (CDM) cosmology,
in which it is (essentially) the growth under self-gravity of 
very tiny initial perturbations in a collisionless fluid 
which give rise to both kind of phenomena. The CMBR fluctuations are 
explained - with apparently great observational success -
by the very tractable evolution of the fluid equations 
linearized in the perturbations. The extension to the 
formation of objects and their distributions is much more 
difficult, involving the highly non-linear regime of the
equations. In practice the main theoretical instrument in
this regime is numerical N-body simulation (NBS)
(\cite{HE}). 
From some given initial conditions (IC - generally a
Gaussian random field with small-amplitude correlated fluctuations) 
specified
by the linear evolution (and thus in principle constrained
by observations of the CMBR) these NBS 
are supposed 
to model the evolution of the continuous density field under 
gravity into the non-linear era. Ultimately catalogues of
different objects are extracted for comparison with 
observations, and the idea is that significant information
about the IC in the continuous density field can be 
extracted by such a comparison. In this letter we re-examine
the dynamics in the non-linear regime of some of the most elaborate
cosmological NBS performed to date, those of the Virgo 
consortium (\cite{virgo}). 
We conclude that the correlations which emerge
at these scales are not correctly understood solely in terms
of the amplification of the initial small fluctuations at large
scales, but rather involve essentially the large fluctuations, 
associated with the discreteness of the simulated system, at the
very smallest scales. We discuss 
the important consequences of this observation.

There are several widely used types of cosmological NBS (\cite{HE}):
`particle-mesh' ($PM$), `particle-particle, particle-mesh' ($P^3M$),
`adaptive mesh refinement' and tree codes. The Virgo consortium's NBS 
are of the adaptive $P^3M$ type, 
and our discussion here applies directly only 
to this type of NBS. The difference between the various methods has to do 
with how the gravitational force acting on a particle is 
modelled. The $PM$ type computes the force from a potential 
calculated on a mesh, typically of the same size as the initial 
mean interparticle 
separation, while the $P^3M$ and other methods add to this the 
direct contribution to the force from particles on scales 
smaller than this distance. 
With either algorithm the relation between the simulated system and
the physical system which one is attempting to model - CDM particles
evolving under their self-gravity - is not at all evident.
In NBS the mass of the `particles' is (very) macroscopic, 
typically of the order of a galaxy mass, and not the 
microscopic mass of a CDM particle. $PM$ NBS 
attempt
to reproduce a treatment of the CDM as a continuous fluid,
with the `particle' putatively representing an element of
the fluid. 
The other methods of NBS 
effectively describe the dynamics of {\it point particles} interacting 
through gravity in an expanding Universe. In particular for $P^3M$
the `particle-mesh' part in the sum of the gravitational force 
can be understood simply 
as a sufficiently accurate and efficient numerical way of 
calculating the force due to points outside the mesh cell, inside 
of which the force is calculated exactly with a smoothed 
$1/r$ potential. This latter smoothing can be understood 
as an effective `smearing' of the particles like in $PM$,
but now at a characteristic scale $\varepsilon$ which,  as we 
will see further below, is always much smaller than the 
mesh size. Physical results are assumed to be those independent
of this smoothing scale, and in fact it can be understood 
as a purely numerical artefact introduced to avoid the 
divergence of the numerical time step in the integration of the 
equations of motions.

In the cosmological context $P^3M$ algorithms have been 
widely used (\cite{virgo,davis85,efs87}) because they allow a much
greater range of spatial resolution while remaining 
numerically feasible. 
Despite its evident importance the central issue of how
this or indeed the $PM$ type of NBS describe CDM has received
surprisingly little attention in the literature. Those few studies 
which have been undertaken of some of these issues 
(\cite{melott90,kms96,smss98}) 
clearly support the idea that discreteness can be very important 
indeed (although possibly less so in $PM$ simulations with a
mesh size well above the interparticle separation).
Our aim here is not to try to resolve this general question.
Rather we are interested in understanding what 
the dynamics is which gives rise to the correlated 
structure in the non-linear regime which is observed 
to emerge in $P^3M$ NBS, 
which as we have remarked 
are simply simulations of self-gravitating particles
subject to some particular initial perturbations.

Usually the main instrument for characterising the evolution
of structure in NBS is the power spectrum 
$P(k) =  |\delta(k)|^2 $ 
where $\delta(k)$ is the Fourier transform (FT) of the density
contrast field. In NBS it is calculated by assigning 
a local mass density on a grid, and then performing a fast FT.
Here our approach is to look 
directly at real space quantities, calculated directly from 
the unsmoothed {\it particle} distributions. Given what we 
have said about the nature of the $P^3M$ NBS, this 
is a reasonable, and, as we will see, instructive approach.
It is the one used  in a recent study (\cite{bottaccio02}) 
through NBS of gravitational dynamics in a purely
Newtonian (non-expanding) system of particles with Poissonian
IC. In this case the method has been used 
to establish the essential role played by 
particle fluctuations 
in the formation of structures.

We have analyzed three cosmological NBS performed
by the Virgo consortium (\cite{virgo}). They are all
CDM models with $N=256^3$ particles and gravitational
force smoothing length $\varepsilon = 0.036  \, Mpc/h$
\footnote{
With this smoothing the force is $53.6 \%$
of the true $1/r^2$ force at $\varepsilon$ 
and more than $99 \%$ at  $2\varepsilon$
(\cite{virgo}).}. 
They all have the same value of the
cosmological parameters, with in particular
total mass density $\Omega = 1$.  The first two
NBS, which we refer to as S-I and S-II,
are of the ``standard'' CDM type, differing
only in their resolution (number density): For S-I  
the volume $V$ of the simulation is a cube of side
$L = 239.5 \, Mpc/h$, while for S-II
$L = 84.5 \, Mpc/h$. The third NBS, S-III,  
is the same as S-I, except that its IC are (slightly) 
different as they are those prescribed by a  
$\tau$CDM model.
In S-I and S-III each particle then represents  
a mass of $m_p= 2.27 \times 10^{11}$ solar masses
while in S-II $m_p = 1.0 \times 10^{10}$. 
In S-I and S-III we find that the initial mean interparticle
separation $ \langle \Lambda_i \rangle  \approx 1 \, Mpc/h$,
while in S-II $\langle \Lambda_i \rangle \approx 0.33 \, Mpc/h$.
Hence $ \langle \Lambda_i \rangle  \gg \varepsilon$ in all three
NBS as noted above. The  NBS are all run from a
red-shift of $z=50$ until today ($z=0$), i.e. for a time
which is essentially the age of the Universe
\footnote{In these matter dominated cosmologies the 
time is given by $t=t_o/(1+z)^{3/2}$ where $t_o$ is the
age of the Universe today.}. This means that the time of 
evolution is essentially one dynamical time
$\tau_{dyn} \approx  1/\sqrt{G\rho}$, where the latter is 
simply the characteristic time scale associated with a 
mass density $\rho= N m_p/V $. A particle
typically moves by a distance of order the lattice spacing 
in this time. 

Let us first consider 
S-I. The first statistic
we consider is the conditional density (\cite{slmp98,hz}) given
by 
$\Gamma(r) = \langle n(r) n(0) \rangle / \langle n \rangle$
where $n(r)$ is the microscopic number density. It is
simply the mean density of points at a distance $r$ from
an occupied point. It is plotted in Fig.\ref{fig1}
for a sequence of time slices, beginning from the 
initial distribution at $z=50$ until today. 
\placefigure{fig1}
Also shown is the pre-initial
`glass' configuration.
This pre-initial distribution (often taken as a regular
lattice in NBS) is obtained by evolving 
the simulation with the sign of gravity reversed, which 
makes the particles settle down to a 
``uniform'' configuration in which the force on each 
particle is very close to zero (\cite{virgo}).  
This distribution is very isotropic but it is still characterized 
by long-range order similar to that of a regular lattice (\cite{hz}). 
It has correlated fluctuations at all scales, and in particular 
large fluctuations (of order one) at scales of order 
$ \langle \Lambda_i \rangle$. The IC of the NBS at the 
initial redshift  $z=50$ are obtained by applying
small (compared to $\langle \Lambda_i \rangle$) displacements
to the points in this  pre-initial distribution, the
displacement field being prescribed by the power spectrum of 
the (continuous) theoretical model  
\footnote{A real space analysis
of these IC  complementary to that 
here is given in (\cite{thierry}). Very significant 
differences between these and those expected
from the theoretical input power spectrum are found
even at large scales.}.
The behaviour of $\Gamma(r)$  for the glassy configuration 
and these IC at $z=50$ are very similar. This is because
the perturbations are very small in amplitude compared 
to the initial interparticle separation $\langle \Lambda_i \rangle$. The 
highly ordered lattice-like 
nature of the initial distribution is manifest:
there is an excluded volume around each point so that
$\Gamma(r)$ is negligible until very close to $\langle \Lambda_i \rangle$,
where it shows a small peak, with some oscillation about
the mean density evident corresponding to the long-range order 
(\cite{hz}).
When the evolution under self-gravity starts the excluded volume 
feature is rapidly diluted, having almost completely disappeared by $z=3$ 
($t=0.125t_o$). This corresponds
to the development of power-law clustering at very small scales where
it was completely absent in the IC.

Let us consider what is driving this dynamics. In
Fig.\ref{fig2} 
we show, for 
the pre-initial and initial perturbed configuration, the mean 
value of the magnitude of the gravitational force 
on a given point due to the mass inside a sphere 
of radius $R$ centred on it. In the pre-initial distribution 
the force is indeed zero within the errors, as expected. 
\placefigure{fig2}
In the
perturbed configurations one sees that the ``long-range'' 
contribution dominates only marginally (by a factor of
about three) over that coming from the nearest neighbour (NN)
(\cite{thierry}).
On the figure is shown also the force on a point due to
a single other one at a distance $\langle \Lambda_i \rangle$: it is, 
as expected,  of order the force due to the nearest points.
When points move 
towards their NN 
the first
contribution grows as one over the distance squared
while the latter changes, supposing that the validity
of linear growth of fluctuations at larger scales,
in proportion to the scale factor, i.e. as $1/(1+z)$.
Thus at $z=5$ we would expect the latter to grow
by a factor of $10$ and thus the NN force to 
dominate below about $\langle \Lambda_i \rangle /5$.
That this is indeed the case is very clearly confirmed 
by considering the evolution of the NN distribution during
the simulation.
In the insert panel of Fig.\ref{fig2} the probability that
the NN of a particle lies at a distance
$r$ is shown, again in different time slices. The initial 
distribution (not shown) is extremely peaked 
about $\langle \Lambda_i \rangle$, then  broadens and 
develops a second peak at small scales. This second
peak at small scales is a clear ``fingerprint'' for the
dominance by NN interactions: it is approximately at 
the scale 2$\varepsilon$ where the force is strongly cut off 
by the smoothing. The build-up
of neighbours around this scale is due to the fact that 
the force which is driving their motion feels the
lower cut-off in it at this scale.

Between $z=5$ and $z=3$ we see - in this range of scales 
completely dominated by NN interactions - the appearance of 
an approximately power-law behaviour in $\Gamma(r)$. At
$z=3$ the amplitude of $\Gamma(r)$ over most of the
range $r <\langle \Lambda_i \rangle $ is greater than 
the mean density so that this power law is also now
seen in the normalised correlation function 
$\xi(r) = \Gamma(r)/ \langle n \rangle -1$
shown in the inset of Fig. 1. At the next time slice, at 
$z=1$, the form of $\Gamma(r)$ up to approximately 
$0.3 \langle \Lambda_i \rangle$ is
almost precisely the same, being simply amplified by an
order of magnitude. At the same time the power law 
extends slightly further before flattening from 
$\Gamma \approx 3  \langle n \rangle$
to a smooth interpolation towards
its asymptotic value. The evolution in the remaining 
two time slices is well described over the range 
$\Gamma(r) > 3 \langle n \rangle$ as a 
simple amplification, and translation towards 
larger scales, of $\Gamma(r)$. At the final time
the power law (with exponent $\gamma \approx -1.7$) 
extends to $\approx 2\langle \Lambda_i \rangle$.

Let us now also consider the other NBS:  S-II
and S-III. 
In both cases we find the same qualitative evolution 
(see Fig.\ref{fig3})
as observed in the S-I simulation in the range of scales
we are discussing. \placefigure{fig3}
The principal difference is in the 
dependence of the amplitude and extent of the power-law
type behaviour in $\Gamma(r)$ as a function of time.  
In S-II the clustering develops more rapidly initially
at small scales, while S-III lags behind relative to S-I.
The reason for this difference is simply traceable to
the relative amplitude of the perturbations in each
case to the pre-initial distribution: S-II has the same
perturbation field superimposed on a distribution with
a smaller initial interparticle separation 
$0.33 \langle \Lambda_i \rangle $, while the perturbation
field at the scale of the lattice in S-III is smaller
in amplitude than in S-I. Thus it takes shorter (or
longer) for the NN contribution to the force to 
dominate and the evolution of clustering due to 
these forces starts faster or slower.

Apart from this difference the evolution is however 
strikingly similar. Note, for example, the almost 
perfect superposition of $\Gamma(r)$ for S-III at
$z=0$ and  S-I at $z=0.3$. The evolution of clustering 
in the non-linear regime thus appears to be almost
independent of the IC,  except 
trivially as  represented by the difference in amplitude. 
Some dependence on the IC 
is evident  at larger scales, but it is a weak residual dependence
imposing a small deviation from the power-law which
is formed at and then translated from the smallest resolved
scales. In Bottaccio et al. (2002)
%\cite{bottaccio02} 
a very similar behaviour 
is observed in the evolution of clustering of particles by
Newtonian forces, without expansion and with
simple Poissonian IC. 
The authors give a physical interpretation of this
clustering in terms of the exportation of the
initial ``granularity'' in the distribution to larger
scales through clustering. The self-similarity in
time of $\Gamma(r)$ is explained as due to a
coarse-graining performed by the dynamics: one 
supposes a NN dynamics between particle-like 
discrete masses, with the mass and physical 
scale changing as a function of time as 
the clustering evolves (particles forming
clusters, clusters forming clusters of clusters
etc.). This would appear to provide a good 
explanation for the behaviour observed 
here as well, but further theoretical 
work is clearly needed to establish this and
find a more quantitative  description.

The primary conclusion we draw from our real space 
analysis of the Virgo NBS is therefore that
the fluctuations at the smallest scales in
these NBS - i.e. those associated 
with the discreteness of the particles -
play a central role in the dynamics of clustering
in the non-linear regime. In particular the 
power-law type correlations appear to be built
up from the initial clustering at the smallest
scales. The nature of the clustering (in
particular the exponent of the power-law) seems 
to be independent (or at most very weakly dependent 
on) the IC  and  its physical origin should therefore
be explained by the dynamics 
of {\it discrete} self-gravitating 
systems. 
The fluid-like statistical description (\cite{hz})  
and  equation of motions, which is the framework used to
describe a CDM universe, 
do not  consider  the non-analytical ``particle'' term  of noise
which is represented by NN interactions. The latter as we have
seen are strongly present in the NBS we have analysed and appear to
play a crucial role in the formation of the correlated
structures observed to emerge.
It is instructive to note that, in the paradigmatic example
of stochastic (homogeneous and isotropic) point processes, the 
Poisson distribution, 
the gravitational force on an average point 
is due very predominantly to by its NN (\cite{chandra43}). 
Large-scale small-amplitude density fluctuations
do not give rise to a significant contribution
to the force acting a point, because of symmetry: i.e. 
large-scale isotropy (\cite{chandra43}).

This conclusion is in disagreement with the usual
interpretation given of the origin of these correlations:
they are supposed to represent {\it solely} the evolution 
in the non-linear regime described by a set of 
fluid equations, with the small fluctuations of
the theoretical input model as IC. 
The dynamics we have described is 
essentially dependent on the gravitational forces at the
smallest resolved scale in the NBS, and 
the small smooth component of the force added to this
by the perturbations at larger scales appears to be
irrelevant in the non-linear regime. What dominates the
dynamics are the large fluctuations at small scales
intrinsic to the discrete pre-initial distribution,
and the imposed fluctuations play only a minor role
in the non-linear regime where the fluctuations are large. 
In cosmology the link between the IC and the ``predictions'' for 
structure formation in the non-linear regime is very important,
as it is through it that one tries to constrain models
using observations of galaxy distributions. Our conclusions
completely change the perspective on this problem. Power-law 
correlations of this type are the most striking and
well established feature of such distributions 
(\cite{pee93,slmp98}). The theoretical problem of
their origin therefore must deal with the apparently crucial
role in their formation of an intrinsically highly fluctuating
(and thus non-analytic) density field. In particular the
origin of the exponent in the correlations and
the dependence of the extent of such correlations 
on the discretization (physical or numerical) needs
to be understood. In particular we note that, despite
the centrality of discreteness, the clustering found
in the Virgo NBS 
(and those of \cite{bottaccio02})
is almost independent of the smoothing scale $\varepsilon$ which
cuts off the force at small distances. We will be 
studying this and other issues in future work.

We thank M. Bottaccio, H. De Vega, R. Durrer, A. Gabrielli, 
A. Melott, M. Montuori, L. Pietronero and N. Sanchez for useful discussions.
F.S.L. acknowledge the support of the Swiss National Science Foundation.

\clearpage

\figcaption[FIG1.eps]
{\label{fig1} Two-point conditional density for the NBS S-I.
The initial mean interparticle separation $\langle \Lambda_i \rangle$
and the softening length $\varepsilon$ are shown, as well
together with the best power-law fit at the end of 
the simulation. In the insert panel the evolution
of the reduced correlation function $\xi(r)$ is shown
by vertical  dashed and dashed-dotted lines respectively. 
The arrows indicate qualitatively how clustering proceeds: 
first it develops at small scales as some particles move 
towards their nearest neighbor, and  then it is subsequentely
exported to larger scales.
}

\figcaption[FIG2.eps]
{\label{fig2} The behaviour of the average total 
force on a point (and its variance) 
due to the mass in a sphere of given radius $r$ about it, for
the pre-initial (glass) and initial particle distribution. The contribution 
from a single particle at distance 
$\langle \Lambda_i \rangle$ is also shown. The units
are given by the choice $Gm_p^2=1$. In the insert panel 
it is shown the evolution of the nearest neighbor distribution 
in S-I. The vertical dashed marks $\langle \Lambda_i \rangle$.}

\figcaption[FIG3.eps]
{\label{fig3}
Comparison of the time evolution of $\Gamma(r)$
in S-I, S-II and S-III, normalized to the mass density.
For S-II and S-III there are shown (in closkwise order) the time slices 
for $z=10,1,0$ while for S-I it is also shown the slice at
$z=0.3$. $\left<\Lambda\right>_{LR}$ is the average distance between
nearest neighbors in the low-resolution simulation (S-I and S-III), 
while  $\left<\Lambda\right>_{HR}$ is that scale 
in the high-resolution simulation
(S-II).} 

%%%UCP%%%
\newpage
\plotone{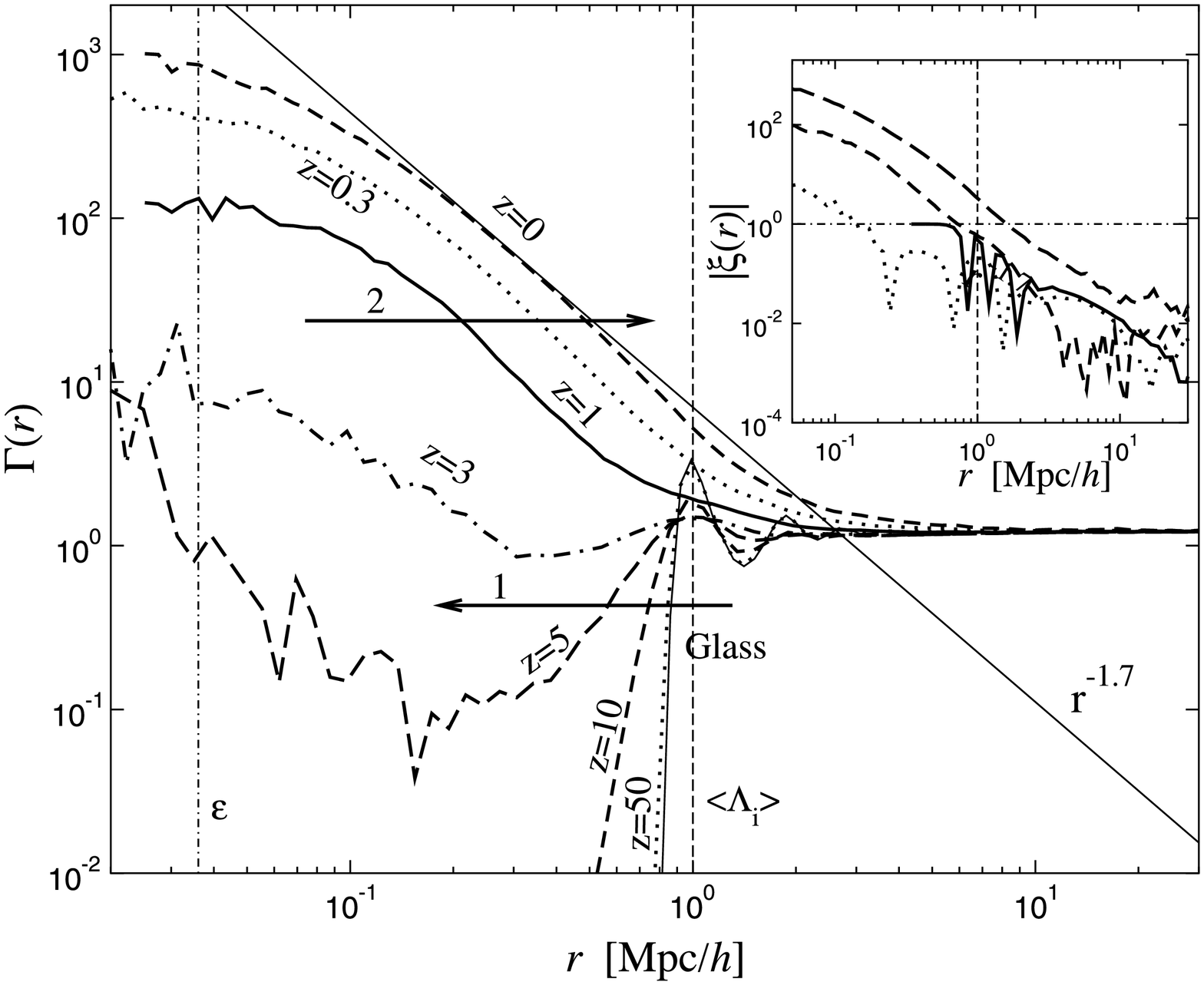}
\newpage
\plotone{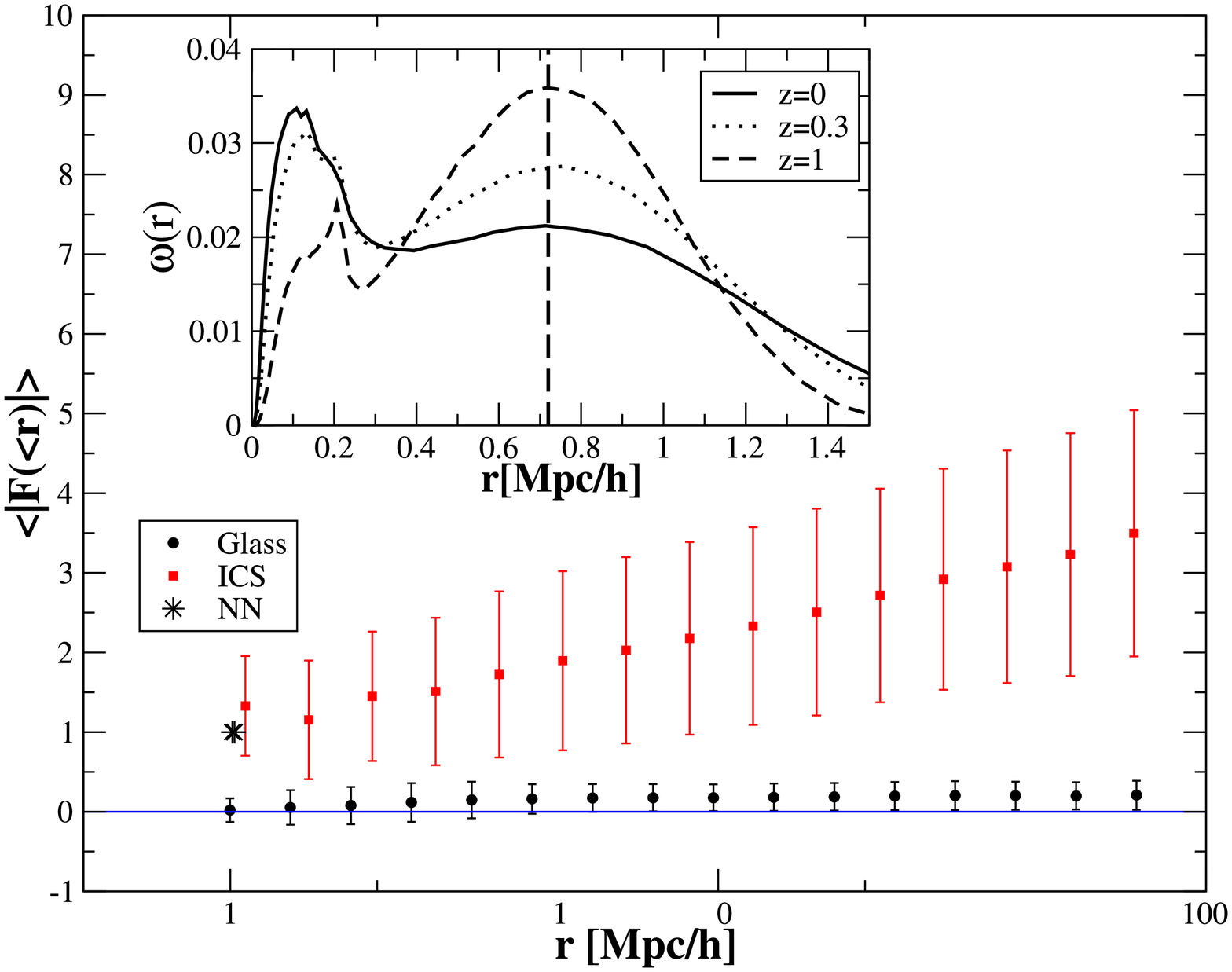}
\newpage
\plotone{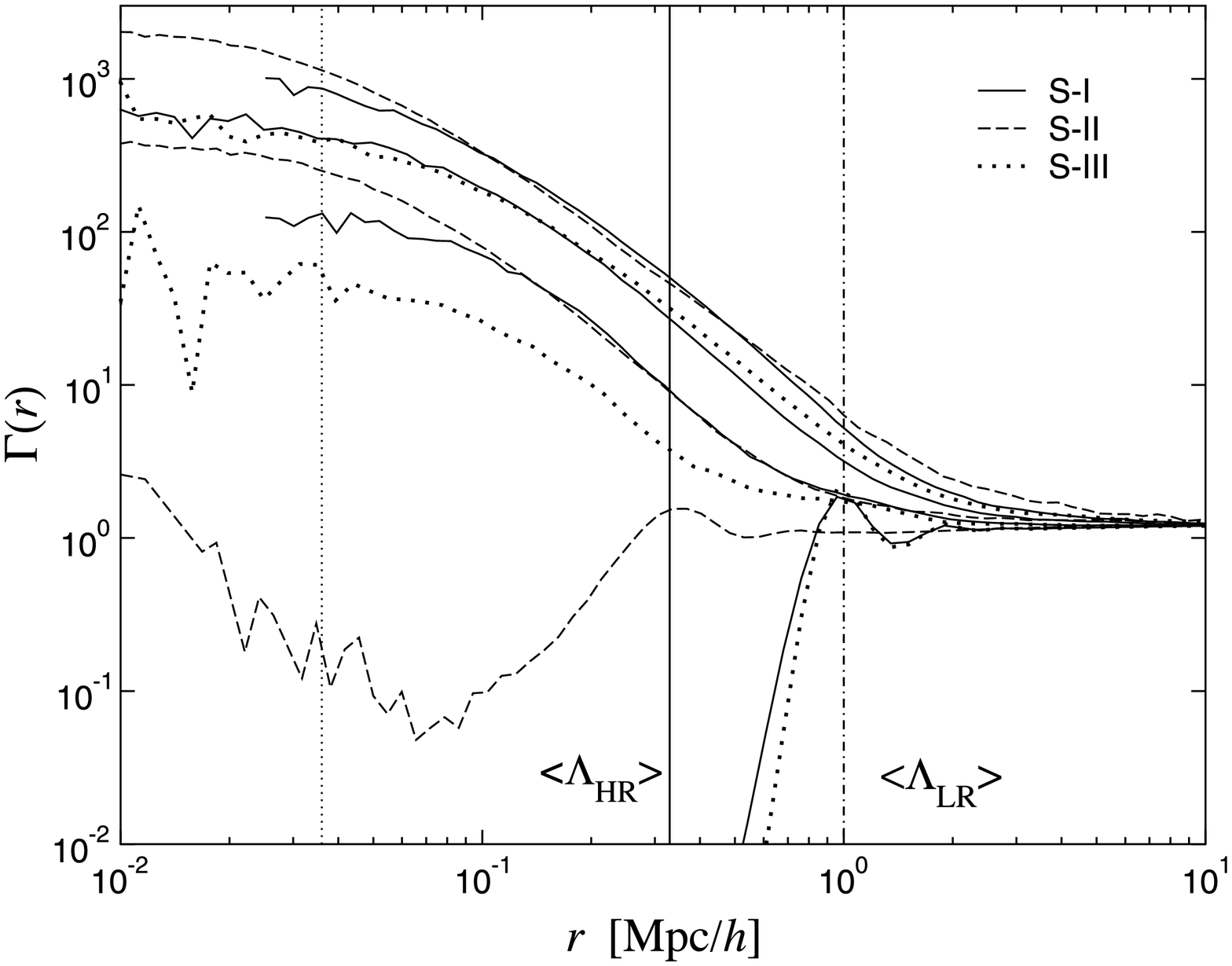}


\begin{thebibliography}{}


\bibitem[Baertschiger \&  Sylos Labini, 2002]{thierry}
 Baertschiger T.  \&  Sylos Labini F.,  2002,
Europhys Lett., 57, 322


\bibitem[Bottaccio et al., 2002]{bottaccio02} 
Bottaccio M. et al., 2002,
Europhys. Lett.,   57, 315 2002

\bibitem[Chandrasekhar, 1943]{chandra43}  
Chandrasekhar S., 1943, 
Rev. Mod. Phys., 15,  1 

\bibitem[Davis et al., 1985]{davis85} Davis M.  et al., 1985,
Astrophys.J.,   292, 371


\bibitem[Efsthathiou and Eastwood, 1981]{ee81} 
Efsthathiou G. and   Eastwood J.W., 1981,
Mon. Not. R. Astron. Soc., 
194, 503


\bibitem[Efsthathiou et al.,  1987]{efs87} Efsthathiou G. et al.,  1987,
Astrophys.J.Suppl.,   57, 241 

\bibitem[Gabrielli, Joyce \& Sylos Labini, 2002]{hz} 
Gabrielli A.,    Joyce M.
and  Sylos Labini F.,  2002,
Phys.Rev., D65, 083523   

\bibitem[Hockney and Eastwood, 1981]{HE}
Hockney R.W. and Eastwood J.W., 1981 Computer simulation using 
particles, McGraw-Hill
%, New York


\bibitem[Khulman, Melott \& Shandarin, 1986]{kms96} Khulman B., Melott A.L.
\& Shandarin S.F., 1986, Astrophys.J.,   470, L41 

\bibitem[Jenkins et al., 1998]{virgo} Jenkins A., et al., 1998, 
Astrophys.J., 499,  20 


\bibitem[Melott, 1990]{melott90} Melott A.L., 1990, 
Comments Astrophys.,  15, 1 


\bibitem[Padmanabhan, 1993]{pad93}  
Padmanabhan T.,  1993, Structure formation in the universe,
Cambridge University Press
%,  Cambridge, England 


\bibitem[Peebles, 1993]{pee93}
Peebles P. J. E.,  1993, 
Principles of Physical Cosmology, Princeton University Press
%, NJ


\bibitem[Splinter et al., 1998]{smss98}  
Splinter R.J.,   Melott A.L.,  Shandarin  S.F. and  Suto Y.,
1998, Astrophys.J.,  497, 38 



\bibitem[Sylos Labini, Montuori  and Pietronero, 1998]{slmp98}
Sylos Labini F., Montuori M., and Pietronero L., 1998, 
Phys.Rep., 293, 66 



\end{thebibliography}
\end{document}